\documentstyle[aps,12pt]{revtex}
\newcommand{\be}{\begin{equation}}
\newcommand{\ee}{\end{equation}}
\newcommand{\ben}{\begin{eqnarray}}
\newcommand{\een}{\end{eqnarray}}

\begin{document}
\title{ An Experiment to Distinguish Between
de Broglie-Bohm and Standard Quantum Mechanics}
\author{Partha Ghose}
\address{S. N. Bose National Centre for Basic Sciences,
JD/III Salt Lake, Calcutta 700 098, India}
\maketitle

\begin{abstract} 
An experiment is suggested that is capable of
distinguishing between the de Broglie-Bohm theory and standard
quantum mechanics.
\end{abstract}
\vskip 0.4in

The proponents of the de Broglie-Bohm quantum theory of motion
(dBB) \cite{Bohm,Bohm2,Holland} have always held that it is
constructed to make exactly the same {\it statistical} predictions
as standard quantum theory (SQT) in every conceivable physical
situation. By statistical predictions is meant averages of
dynamical variables over all possible ``hidden'' trajectories at a
fixed instant of time (a virtual Gibbs ensemble). The average over
a Gibbs ensemble usually turns out to be the same as the average
over a time ensemble, i.e., an ensemble built over clearly
separated short intervals of time. Although this equivalence holds
in SQT, I will show that it breaks down for a special two-particle
system in dBB when a certain combination of bosonic and
geometrical symmetries holds. It is possible to exploit this
feature and do an experiment with a sequence of photon pairs that
can distinguish between SQT and dBB, so far thought to be
statistically completely equivalent.

To see how predictions of dBB and SQT can defer in a special case, let
us  consider a variant of the double-slit experiment
in which the wave packets of a pair of identical
non-relativistic particles $1$ and $2$ are simultaneously
diffracted by two slits $A$ and $B$ (which have non-zero widths $d$
much larger than the de Broglie wavelength of the packets, and
can be regarded as the sources of
the diffracted waves) and overlap in some region $\cal{R}$ (see figure)
sufficiently
far from the slits so that the waves in that region can be
regarded as plane waves (the case of Fraunhoffer diffraction).
Let us assume that the particles are bosons. The problem is essentially a
two dimensional one. So let this plane be the $y-x$ plane, and let us
choose the origin $O$ of the coordinate system to be at the mid-point of the
line joining the two slits $A$ and $B$ as shown in the figure.
Then the two-particle wave function in the region $\cal{R}$ in the
$x-y$ plane can be written as (\cite{Holland})

\ben \Psi (\vec{r}_1, \vec{r}_2, t ) = \sqrt{N} g [
e^{i(\vec{k}_A.\vec{r}_1 + \vec{k}_B.\vec{r}_2)} +
e^{i(\vec{k}_A.\vec{r}_2 + \vec{k}_B.\vec{r}_1)}] e^{-iEt}
\label{eq:1} \een where $N$ is a normalization factor, $g$ is a
gaussian factor that takes account of the widths of the two slits,
and the $k$'s are wave vectors (with $\vert \vec{k}_A\vert = \vert
\vec{k}_B\vert$). Since the double-slit arrangement has a natural
symmetry about the line $x = 0$, this wave function must be
symmetric both with respect to the exchange of the two particles
and reflection in the line $x = 0$. It is straightforward to show
that this wave function can be written in the form

\ben \Psi(\vec{R}, \vec{r}) = \sqrt{N} g e^{i \vec{K}.\vec{R}}
[e^{i \vec{k}.\vec{r}} + e^{-i \vec{k}.\vec{r}}] \een where

\ben \vec{K} &=& \vec{k}_A + \vec{k}_B \\ \nonumber \vec{k} &=&
\vec{k}_A - \vec{k}_B\\ \nonumber \vec{R} &=& \frac{1}{2}
(\vec{r}_1 + \vec{r}_2)\\\nonumber \vec{r} &=& (\vec{r}_1 -
\vec{r}_2) \een The vector $\vec{K}$ is the centre-of-mass
momentum of the two-particle system and $\vec{R}$ is the radius
vector of the centre-of-mass. The symmetry of the arrangement
guarantees that the $x$-component of $\vec{K}$ is 
zero ($K_x = 0$) (and hence $k_y = 0$). (Any $\vec{K}$ and $\vec{k}$ which have
these properties are allowed.) It is clear from all this that the
relative motion of the particles is decoupled from the
centre-of-mass motion. This is a consequence of translation
symmetry which holds with plane waves. It is easy to see from (2)
that

\ben \Psi^* \Psi &=& 2 g^2 N [ 1 + cos 2 (\vec{k}. \vec{r})]\\
\nonumber &=& 2 g^2 N [ 1 + cos \frac{2 \pi (x_1 - x_2)}{L})] \een
where $L\approx \lambda/2 \theta$ is the classical fringe spacing.
The probability of joint detection of the particles around points
$x(P)$ and $x(Q)$ on a screen with fixed $y$ in the region
$\cal{R}$ is therefore given by

\be
P_{1 2} \biggl(x (P), x(Q)\biggr) \delta x(P) \delta x(Q) =
\int_{x(P)}^{x(P) + \Delta x (P)} d r_1
\int_{x(Q)}^{x(Q) + \Delta x(Q)} d r_2 \vert \Psi \vert^{2}
\label{eq:2}
\ee
which is to be evaluated on a $t=$ constant spatial surface. It
contains fourth-order interference terms between $1$ and $2$
\cite{Ghosh}, \cite{Horne}. (The small but finite domains of
integration are chosen to take account of the finite size of the detectors.)

So far we have used only standard quantum mechanics (SQT). Let us now
introduce the guidance condition of dBB and calculate the Bohmian
velocities of particles $1$ and $2$. They are given by

\ben
\vec{v}_1 &=& \frac{\vec{j}_1}{\Psi^* \Psi} =
\frac{\hbar}{2 m}  \vec{K} \\ \nonumber
\vec{v}_2 &=& \frac{\vec{j}_2}{\Psi^* \Psi} = \frac{\hbar}{2 m}
\vec{K} 
\label{eq:3}
\een
where $\vec{j}_1$ and $\vec{j}_2$ are the convection currents of the
two particles. It follows from this and the fact that $K_x = 0$ that
$v_{1x}=v_{2x}=0$ and so
\ben
v_{1x} + v_{2x} = 0
\een
This implies that

\ben
x_1(t) + x_2(t) = x_1(0) + x_2(0)
\een
Thus, if the initial positions of the two particles are 
symmetrical about the
line of symmetry ($x = 0$), i.e., if $x_1(0) + x_2(0) = 0$, we must have

\ben
x_1(t) + x_2(t) = 0
\label{eq:5}
\een
for all times, i.e., the trajectories will always be symmetrical and
parallel about
this line. This is the main source of incompatibility between dBB
and SQT.

It is important to draw attention to the crucial role played by the
combination of bosonic and geometric symmetry in the above argument.
Notice that if bosonic symmetry is not imposed on the two-particle
wave function (equation (\ref{eq:1})), it will not be symmetric under
reflection in the line $x=0$. This will mean that with Maxwell-Boltzmann
statistics (only one term in equation (\ref{eq:1})), for example, the
trajectories of the two particles can cross each other and the line $x=0$.
But the situation is different with bosonic symmetry. This additional
symmetry constrains the individual particle trajectories
not to touch each other and this line
\cite{Holland}. This brings about a fundamental change---the
particle trajectories are clearly separated into two
non-intersecting disjoint classes, one above and one below the line
of symmetry.

To see what this implies for the two-particle double-slit experiment
under consideration, consider the dBB ensemble to be built up of single
pairs of particle trajectories arriving at the screen at different instants
of time $t_i$ such that the joint probability of detection is given by

\ben
P_{1 2}
= lim_{N \rightarrow \infty}\sum_{i=1}^N
\frac{1}{\delta (0)}\int d x_1 \int d x_2 P(x_1,x_2,t_i)\nonumber\\
\delta (x_1 - x_1(t_i))\,\delta (x_2 - x_2(t_i)) \delta (x_1(t_i) + x_2(t_i))\nonumber\\
= lim_{N \rightarrow \infty} \sum_{i=1}^N P(x_1(t_i), -x_1(t_i)) = 1
\label{eq:9}
\een
where the constraint (\ref{eq:5}) has been taken into account. Every
term in the sum represents only one pair of trajectories arriving at
the screen at the points $(x_1(t_i), -x_1(t_i))$ at time $t_i$,
weighted by the corresponding density $P$. In discrete time ensembles
of this type every pair can be separately identified, and it is clear
that if the detectors are placed symmetrically about the plane $x=0$,
they will record coincidence counts just as predicted by SQT. On the
other hand, if they are placed asymmetrically about $x=0$, the
joint detection of every pair, and hence also their time average,
will produce a null result which is {\it in conflict with the
SQT prediction} \cite{footnote}.

If all the times $t_i$ are put equal to a fixed time $t$ in (\ref{eq:9}),
the sum over $i$ can be converted to an integral over all the trajectories
which pass through all the points of the screen at that time
(a Gibbs ensemble). Now consider the joint probability of detection
around two points $x(P)$ and $x(Q)$ on the screen that are not symmetric
about $x=0$. The trajectories that pass through these regions cannot be
partners of individual pairs which are constrained by (\ref{eq:5}),
and must therefore belong to {\it different} pairs which are
{\it not so constrained}. If the points are symmetrically situated
about $x=0$, the constraint is automatically satisfied. The joint
probability of detection in this case is therefore given by

\ben
P_{12}\biggl(x(P),x(Q),t\biggr)
= \int_{x(P)}^{x(P) + \Delta x (P)} d x_1(t)
\int_{x(Q)}^{x(Q) + \Delta x(Q)} d x_2(t)\nonumber\\
P(x_1(t), x_2(t))
\een
which is, in fact, the same as the SQT prediction (\ref{eq:2}).
It becomes impossible therefore to
distinguishing between dBB and SQT using such an experiment, and the
trajectories remain ``hidden'', as
correctly maintained by Bohm.

Exactly the same conclusion can be drawn for a pair of photons
by using a consistent quantum mechanical formalism for massless
relativistic bosons below the threshold of pair creation \cite{Ghose}.
In an actual experiment with photons,
one must choose identical photons generated by degenerate parametric
down-conversion, for example, and make the signal and idler photons pass
through the two slits $A$ and $B$ simultaneously. One must also cut down
the intensities of the beams to the single pair level. To the best of
my knowledge all the conditions necessary to show incompatibility between dBB 
and SQT have not been met with in any experiment
done so far \cite{Ghosh}. Hence the necessity for a critical experiment
which can settle the fundamental questions as to whether the wave function
description is the most complete possible and also whether the lack of
causality in quantum phenomena is really fundamental.

I am grateful to Franco Selleri, Rupa Ghosh,
S. M. Roy, C. S. Unnikrishnan, A. S. Majumdar and B. Dutta-Roy for many
helpful
discussions. I also wish to acknowledge financial support from the
Department of Science and Technology, Government of India, through a
research grant.

\vskip 0.2in

FIGURE CAPTION: Two-particle double-slit experiment

\pagebreak

\setlength{\unitlength}{1mm}
\begin{picture}(120,100)
\put(1,0){\line(1,0){105}}
\put(105,0){\vector(1,0){18}}
\put(10,0){\line(0,1){15}}
\put(10,0){\line(0,-1){15}}
\put(10,-12){\line(0,-1){10}}
\put(10,12){\line(0,1){10}}
\put(10,25){\line(1,1){70}}
\put(10,24){\line(1,-1){70}}
\put(10,-24){\line(1,1){70}}
\put(10,-25){\line(1,-1){70}}
\put(10,27){\vector(0,1){18}}
\put(10,-27){\vector(0,-1){18}}
\put(80,0){\line(0,1){100}}
\put(80,0){\line(0,-1){100}}
\put(67,4){\LARGE $\cal{R}$}
\put(5,45){x}
\put(7,1){0}
\put(120,-3){y}
\put(5,23){A}
\put(5,-25){B}
\put(83,10){P}
\put(83,-23){Q}
\put(80,10){-}
\put(80,-18){-}

\end{picture}


\vskip 0.2in
\begin{thebibliography}{99}

\bibitem{Bohm}
D. Bohm, {\it Phys. Rev.} {\bf 85}, 166, (1952).

\bibitem{Bohm2}
D. Bohm and B. J. Hiley, {\it The Undivided Universe}
(Routlege and Chapman $\&$ Hall, London, 1993).

\bibitem{Holland}
P. R. Holland, {\it Quantum Theory of Motion} (Cambridge
University Press, 1993).

\bibitem{Leavens}
C. R. Leavens, {Found. of Physics} {\bf 25}, 229, (1995).

\bibitem{Pais}
A. Pais, {\it Subtle is the Lord...} (Oxford University Press, Oxford,
1982).


\bibitem{Ghosh}
R. Ghosh and L. Mandel, {\it Phys. Rev. Letters} {\bf 59}, 1903, (1987).

\bibitem{Horne}
M. A. Horne, A. Shimony and A. Zeilinger, {\it Phys. Rev. Letters}
{\bf 19}, 2209, (1989);\\
G. Jaeger, M. A. Horne, A. Shimony, {\it Phys. Rev. A} {\bf 48}, 1023, (1993).

\bibitem{footnote}
Implicit in this is the necessary requirement in dBB that only particles
produce detections and not `empty' parts of wave functions. See Ref. 3.

\bibitem{Ghose}
P. Ghose, {\it Found. of Physics} {\bf 26}, 1441, (1996).
\end{thebibliography}
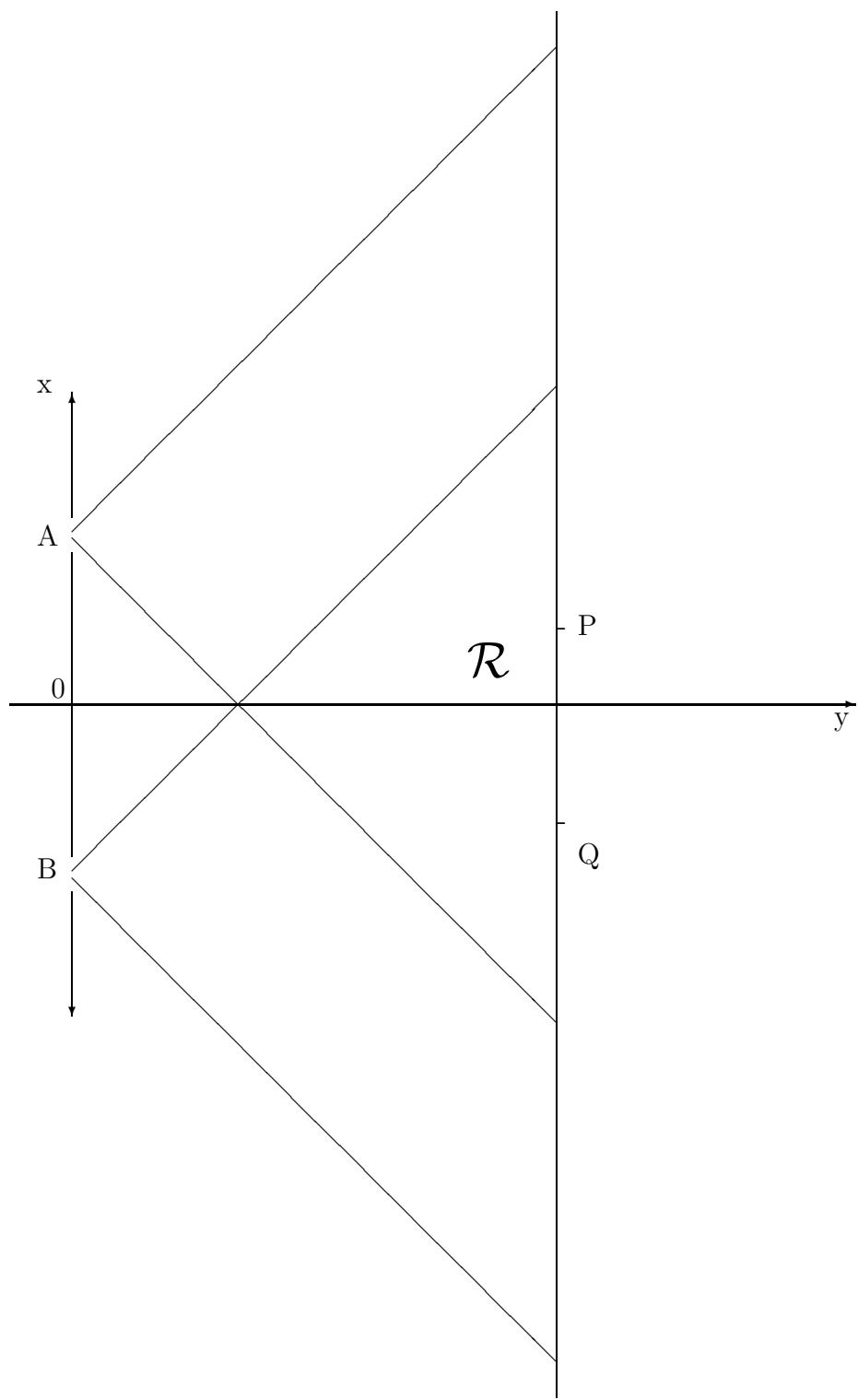
\end{document}